\documentclass{edp-conf}
\usepackage{graphicx}
\begin{document}

\TitreGlobal{SF2A 2004}

\title{Virtual Instrumentation to study galaxies}
\address{GEPI-Observatoire de Paris, UMR8111}
\address{Observatoire de Lyon, UMR142}
\address{Sternberg Institute, Moscow State University}
\address{Observatoire de Gen\`eve, Sauverny, Switzerland}
\author{Prugniel, Ph. $^{1,2}$}
\author{Chilingarian, I. $^3$}
\author{Flores, H. $^1$}
\author{Guibert, J. $^1$}
\author{Haigron, R. $^1$}
\author{J\'egouzo, I. $^1$}
\author{Royer, F. $^{1,4}$}
\author{Tajahmady, F. $^1$}
\author{Theureau, G. $^1$}
\author{V\'etois, J. $^1$}
\runningtitle{MIGALE}
\setcounter{page}{237}
\index{Prugniel, Ph.}
\index{Chilingarian, I.}
\index{Flores, H.}
\index{Guibert, J.}
\index{Haigron, R.}
\index{J\'egouzo, I.}
\index{Royer, F.}
\index{Tajahmady, F.}
\index{Theureau, G.}
\index{V\'etois, J.}

\maketitle
\begin{abstract} 
The MIGALE project provides databases and data analysis tools
to study the evolution of galaxies from z=1 to z=0. It develops
and maintain a general database, HyperLeda, to give a homogenized
parameterization for 3 million objects, and several archives or
specialized databases. It also develops tools to analyse on-the-fly data
extracted from the database or obtained through the Virtual Observatory
(Virtual Instruments). The package made for this project, Pleinpot,
is distributed as open source.

\end{abstract}
%

\section{How the Virtual Observatory approach can help for the data analysis?}

The goal of the Virtual Observatory (VO, see http://www.ivoa.net/)
is to enhance the scientific efficiency by providing a transparent access
to the archive data stored throughout the world, so that the end user will 
have the filling that all these resources are located on its own computer. 
To make this enormous quantity of data useful (volumes are counted in Pbytes),
high performance discovery tools are needed, and a condition is to provide 
an accurate description of any individual dataset.

A precise description also enables to change the strategy
for the data analysis. For example, it will not be required anymore
to tell to the analysis program what is the spectral resolution and
how it changes with wavelength: This information will come with the data,
as part of the so-called {\it meta-data} or more familiarly FITS keywords.
It will be possible soon to push one step forward the automation
of the data analysis.

Why is it desirable to automate analysis pipelines? Naturally, the main
reason is that it is expected to be faster, and possibly more reliable
since it eliminates some risks of errors during the preparation of the
reduction procedure. The complexity of data analysis also increases
as it allows to put finer constraints to the details of the physical
mechanisms. So, astronomers interested in a specific phenomenon and using
multi-wavelength data can hardly be experts in all the aspects of
the data analysis; more automatic procedures are simpler to use and
enable non-specialists to interpret the data. There is obviously the risk
to miss-use a procedure, as for example give a wrong combination
of parameters to a model, but astronomers are highly aware of this
danger which anyway already exists when we are using the present software.
The VO does certainly not worsen the situation.

Actually, not only it is desirable to automate processing, but it is
necessary. Some spectrographs produce millions of spectra over their
lifetime, and it is absolutely impossible to make a detailed interactive
analysis of each of them (see for example the analysis of SDSS data
in Mathis et al. 2004).

Analysis pipelines, to compare models to observations, will be developed
either as client packages to be run on the user's machine or as online
services run on a remote server fed by data provided by the user or
extracted from the VO and returning results in the
formats defined by the VO so they can be used by other VO tools.
Such services already exist. For example, a VO implementation of sextractor was
used at the 2004 AVO science demo (Padovani et al. 2004).

\section{The MIGALE project}

The MIGALE project (http://www.sai.msu.su/migale/) 
has been started to share common developments
between different scientific projects aimed at studying the evolution
of galaxies from medium to low red-shifts.

MIGALE proposes some databases. In particular HyperLeda is a general
database for studying the physics of nearby galaxies. It contains about
3 millions objects for which data are collected from the literature and
from the surveys and are homogenized to a common scale. Data
concerns the structure, kinematics, stellar and gas content of galaxies.
They are either integrated measurements or spatially resolved information,
as photometric or kinematical profiles. HyperLeda stores also an
archive of reference FITS data, images or spectra, and offers some
prototype of Virtual Instruments, such as the on-line version of
P\'EGASE (Le Borgne et al. 2004).

Other new databases made in the frame of the MIGALE project are presented
at this conference. HiGi (Theureau et al. 2004) describes the HI content of galaxies
and gives access to HI raw and processed observations.
The Giraffe Archive (Royer et al. 2004) delivers processed data from 
the Flames/Giraffe instrument at ESO. 
ASPID (Chilingarian et al. 2004) is archive 
containing the raw data from the Russian 6 m telescope.

The different developments made in the frame of MIGALE share the
same software package, Pleinpot 
(http://leda.univ-lyon1.fr/pleinpot/pleinpot.html)
providing in particular a library for database access, retrieval and
formatting of the data and for data processing and analysis. The
software is freely available under the terms of the GNU Public License.
Pleinpot is used in other projects, as the ELODIE-SOPHIE archive
from Observatoire de Haute-Provence (Moultaka et al. 2004).

MIGALE results from a long term effort initiated by the Programme
National Galaxies (CNRS) to coordinate the initiatives started in different
laboratories, in particular in Observatoire de Lyon and in GEPI (Observatoire
de Paris).

\section{The Virtual Instruments planned by MIGALE}

The main focus of MIGALE is on integral field spectroscopy. A large program 
is conducted using the Giraffe
spectrograh at the ESO/VLT to study the gas content, kinematics and
environment of distant galaxies. First results of velocity fields
are presented in Flores et al. (2004) and the deconvolution method
is presented in Puech et al. (2004). The corresponding package, Disgal3D,
will evolve toward a Virtual Instrument using inputs from high
spatial resolution images (HST) to deconvolve lower resolution data cubes
from ground based integral field units.

In the Local Universe we are studying the counterparts of the star forming
galaxies observed at larger distances. We are using the Russian
6 m telescope equipped with the MPFS integral field spectrograph to
observe galaxies from the Virgo cluster. We are in particular interested
in the smaller galaxies, such as diffuse ellipticals (dE), which have
apparently formed their stars about 5 to 7 Gyr ago. A preliminary 
scientific result presented in Chilingarian and Prugniel (2004) shows a 
complex kinematical structure in a dE
and demonstrates our ability to constraint the history of the stellar
population.
\\

\begin{figure}[h]
   \centering
   \includegraphics[width=8cm]{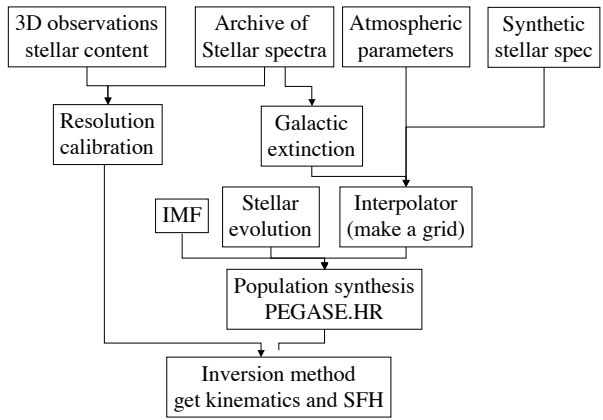}
      \caption{ Architecture of the Virtual Instrument to study the
stellar population of galaxies.}
       \label{figure_mafig}
   \end{figure}

The general principle to study a stellar population from a spectrum
integrating all the light along a line-of-sight is to compare
the observation with the spectrum of a modeled stellar population
convolved by the appropriate distribution of internal velocities.
Fig. 1 presents the main modules intervening in this process.

The synthetic spectrum may be either based on theoretical or empirical 
stellar spectra, eg. Prugniel \& Soubiran
(2004). Empirical spectra need to be corrected for Galactic extinction
and to be used to produce an interpolated grid, the atmospheric
parameters must be available, for example from Soubiran et al (2002).
The population synthesis package is fed by this grid 
and needs also  models for the stellar evolution (evolutionary tracks),
for the initial mass function and scenario of enrichment and star formation.
A new package, P\'EGASE.HR, has recently been published
(Le Borgne et al. 2004).
The analysis of 1 or 2-D spectra then requires to calibrate
the spectral resolution and the photometry using an archive of stellar
spectra and then to use elaborated inversion methods, as eg. described
in Ocvirk et al. (2004).

It is clear that the various modules enumerated here have a broader
interest than this specific application. This calls for a separate
implementation of these elementary bricks, the VO protocols and formats
being used to interconnect them easily.

Though it is possible to execute all these modules on the fly, using
the very last version of the stellar libray, extinction model, ...,
this would be heavily demanding in CPU resources and
would introduce a large number of free parameters whose incidence on the
result is difficult to assess.
For these reasons, we are adopting compromises. 
In particular, the grids of stellar spectra used for synthesis
are stored statically and updated only when either the archive, the reduction
procedure or calibration is significantly improved.

Presently, this Virtual Instrument to study the stellar populations is
still under development and gives its first scientific results. We expect
to deliver a robust implementation of a tool to retrieve the internal
kinematics and the stellar formation history within 3 years. It
will be available both as an online service and as a client package
that can be installed on the user's machine.


\end{document}